\documentclass[format=sigconf, review=false, anonymous=false, screen, dvipsnames]{lib-acm/acmart}

\setcopyright{acmlicensed}
\copyrightyear{2018}
\acmYear{2018}
\acmDOI{XXXXXXX.XXXXXXX}

\copyrightyear{2026}
\acmYear{2026}
\setcopyright{cc}
\setcctype{by}
\acmConference[CHI '26]{Proceedings of the 2026 CHI Conference on Human Factors in Computing Systems}{April 13--17, 2026}{Barcelona, Spain}
\acmBooktitle{Proceedings of the 2026 CHI Conference on Human Factors in Computing Systems (CHI '26), April 13--17, 2026, Barcelona, Spain}
\acmPrice{}
\acmDOI{10.1145/3772318.3790786}
\acmISBN{979-8-4007-2278-3/2026/04}

\usepackage{booktabs} %

\usepackage{exii-macros}

\usepackage{booktabs}

\hidecomments

\makeatletter
\g@addto@macro\normalsize{%
  \setlength\abovedisplayshortskip{-9pt}
  \setlength\belowdisplayshortskip{3pt}
}
\makeatother

\begin{document}

\tolerance=400 

\title[AI Margin Notes]{Designing and Evaluating AI Margin Notes in Document Reader Software}

\author{Nikhita Joshi}
\orcid{0000-0001-9493-7926}
\affiliation{%
  \institution{Cheriton School of Computer Science\\University of Waterloo}
  \city{Waterloo, Ontario}
  \country{Canada}
}
\additionalaffiliation{%
  \institution{LISN, Universit\'{e} Paris-Saclay, CNRS, Inria}
\city{Orsay}
  \country{France}
}
\email{nvjoshi@uwaterloo.ca}

\author{Daniel Vogel}
\orcid{0000-0001-7620-0541}
\affiliation{%
  \institution{Cheriton School of Computer Science\\University of Waterloo}
    \city{Waterloo, Ontario}
  \country{Canada}
}
\email{dvogel@uwaterloo.ca}

\renewcommand{\shortauthors}{Joshi and Vogel}

\begin{abstract}
AI capabilities for document reader software are usually presented in separate chat interfaces. We explore integrating AI into document comments, a concept we formalize as AI margin notes. Three design parameters characterize this approach: margin notes are integrated with the text while chat interfaces are not; selecting text for a margin note can be automated through AI or manual; and the generation of a margin note can involve AI to various degrees. Two experiments investigate integration and selection automation, with results showing participants prefer integrated AI margin notes and manual selection. A third experiment explores human and AI involvement through six alternative techniques. Techniques with less AI involvement resulted in more psychological ownership, but faster and less effortful designs were generally preferred. Surprisingly, the degree of AI involvement had no measurable effect on reading comprehension. Our work shows that AI margin notes are desirable and contributes implications for their design.

\end{abstract}



\begin{teaserfigure}
  \includegraphics[width=\textwidth]{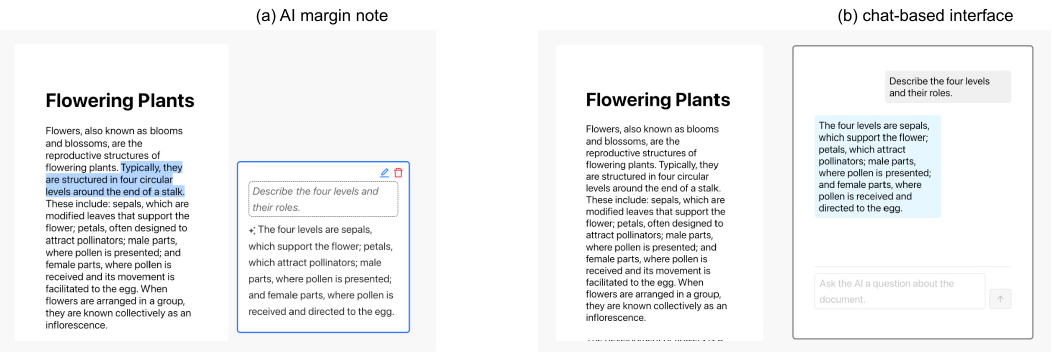}
  \caption{Different ways to leverage LLM capabilities in document reader software: (a) we propose AI margin notes that are integrated with the document text unlike (b) chat-based interfaces that are separated from the document text.}
  \label{fig:teaser}
\end{teaserfigure}

\maketitle

\section{Introduction}
Taking notes while reading documents is a common active reading strategy that can be done in two primary ways. First, people can write notes that are decoupled from the document text by writing on a separate piece of paper. Second, people can write notes that are integrated with the document text \cite{AdlerDiaryStudy}, for example, by writing in the margins (\emph{marginalia}). Such `margin notes' have been common practice for centuries, as they provide a space for readers to summarize, paraphrase, explain unfamiliar concepts, make connections to existing knowledge, and even express personal opinions \cite{jackson2001marginalia, porter2004beyond}. This can be especially beneficial when reading for educational purposes \cite{bold2017marginalia}, as it provides space for people to work through difficult sections and quickly re-access their thoughts afterwards within a shared context \cite{marshall1997annotation, AdlerDiaryStudy}.
With digital documents, margin notes can be added as digital ink with a pen (e.g., \cite{romat2019spaceink}), but a more common way is leaving \emph{comments} when reading documents in systems like Google Docs, Microsoft Word, and Adobe Acrobat Reader \cite{wolfe2001margins}. This is usually done by selecting text within a document and writing in a text box positioned beside it.

Large language models (LLMs) can support reading documents to summarize, reword, and better understand unfamiliar concepts \cite{GoogleGuidroz2025llm, GuTextSmimming2024, kreijkes2025effects}. The generated output serves a purpose like margin notes, yet the capabilities of LLMs have not been integrated into commenting features of document reader software. Instead, most LLM-enhanced document readers use a separate, chat-like interface disconnected from the document. There is an opportunity to improve how LLMs are used in this context by integrating their capabilities directly into `margin note' comments that are linked to the document text.

We explore the design of ``AI margin notes'' that leverage LLMs to enhance comments in document reader software through three controlled experiments. Every experiment required participants to read short, non-fiction documents while interacting with an LLM, primarily within comments, and answer reading comprehension questions two hours later. Each experiment focused on a different design parameter. First, we compared AI margin notes to traditional, chat-based prompting to better understand the effects of \emph{integration}. Second, we compared manually selecting text that an AI margin note is associated with, to automatically selecting text through an LLM-powered assistant to better understand the effects of \emph{selection automation}. Third, we explored different AI margin note techniques that leverage LLMs in different ways to understand the effects of \emph{human and AI involvement}. Specifically, AI margin notes that generate summaries, fill-in-the-blank exercises, responses to specific prompts, practice short answer questions, and feedback on how written text can be improved.

Our results indicated that AI margin notes were preferred over chat interfaces and that selecting text for AI margin notes should be done manually. AI margin note techniques with varying levels of human and AI involvement were valued for different reasons. For example, techniques with more human involvement were typically associated with more psychological ownership, but techniques with more AI involvement were faster, less effortful, and generally preferred. 
Our work contributes:
\begin{itemize}
    \item the idea of AI margin notes: comments that are enhanced with LLMs to support note-taking that is integrated with the document text; and
    \item empirical results from three experiments, each focusing on a specific design parameter, demonstrating that AI margin notes are a desirable feature for document reader software.
\end{itemize}

\section{Background and Related Work}
AI margin notes relate to existing literature in psychology about the benefits and challenges of note-taking while reading, and other techniques that have used LLMs to improve reading and note-taking. We focus specifically on relevant work focused on reading, rather than taking notes while attending a lecture or watching a video.

\subsection{Note-Taking while Reading}
There are many types of notes that readers can create, margin notes being one of them. 
Research in psychology suggests that the general activity of note-taking is beneficial for two main reasons \cite{KiewraNoteTakingReview}. First, notes act as external storage for information, which can be re-read to reinforce memory (the \emph{storage function}). Second, the act of creating notes can facilitate learning as it requires readers to pay more attention to the material to process and organize ideas (the \emph{encoding function}). 

The encoding function is especially beneficial when the reader processes the text at a deeper level \cite{kiewra1985investigating, craik1972levels}, for example, by connecting it to prior knowledge and experiences. However, many people opt for shallower and less effective note-taking strategies. For example, Bretzing and Kulhavy's analysis of students' note-taking activities \cite{bretzing1981note} revealed that students tend to take verbatim notes that repeats text from documents. In another study \cite{bretzing1979processing}, they showed that students who took verbatim notes performed worse on a test than those who used deeper note-taking strategies by writing their own summaries or by paraphrasing text from the document.

Note-taking is a complex activity that requires significant cognitive effort as readers must coordinate and frequently switch between reading and writing activities. Writing takes more time than reading, and excessive delays between reading activities as a result of note-taking can hinder comprehension. Therefore, readers often experience significant mental and temporal demand, even when they are reading and taking notes without any time limits \cite{piolat2005cognitive}. This can become especially tiring with longer \cite{kobayashi2005limits} or poorly-formatted documents \cite{olive2017processing}. Although researchers have argued that increased cognitive effort can benefit learning \cite{bjork1994memory, bjork2011making, bjork1994institutional} and memory \cite{tyler1979cognitive}, it does not always lead to improved learning outcomes, for example, if the task is too frustrating or if the learner lacks motivation \cite{grund2024learning}.

\subsection{Reading and Note-Taking with LLMs}
Recent work has investigated how LLMs can improve user experience and comprehension while reading and taking notes. For example, text simplification techniques that turn complex documents into simplified versions \cite{GoogleGuidroz2025llm, AugustPaperPlain2023} and make documents easier to skim \cite{GuTextSmimming2024} can improve reading comprehension and reduce workload.
Users of systems like ChatGPT, Adobe Acrobat \cite{AdobeAcrobatAI}, Google Notebook LM \cite{notebooklmGoogleNotebookLM}, NoteGPT \cite{NoteGPT}, and ChatPDF \cite{chatpdfChatPDFChat} can ask questions, summarize, and write notes about documents. However, few investigations examine the effect on factors like reading comprehension, reading duration, workload, and preferences.

Kreijkes et al. \cite{kreijkes2025effects} asked high school students to try different note-taking techniques to study for reading comprehension tests that took place three days later: note-taking independently and note-taking while having access to an LLM-powered chatbot to ask questions. Both of these note-taking techniques were compared to a baseline of asking the chatbot questions about the document, without any note-taking. Their results showed that both note-taking techniques led to better performance than just asking the chatbot questions about the document, suggesting that using LLMs in a more cognitively engaging way (i.e., with some note-taking) can improve reading comprehension. However, they did not compare the two note-taking conditions, making the effect of LLM use on note-taking unclear.

Although Kreijkes et al. found that just asking the LLM questions led to poorer test performance, it was preferred by participants as it was perceived to be more enjoyable and less effortful. Such findings have been reproduced in other studies
focused on the effects of LLMs on learning more broadly. Systems like ChatGPT can allow for personalized learning experiences, which can improve academic performance, reduce mental effort, and motivate learners \cite{wang2025effect, deng2025chatGPT}, however, learners may over-rely on these systems \cite{wang2025effect}. A balance may be to encourage more cognitive engagement. For example, ChatPRCS \cite{wang2024chatprcs} generates practice reading comprehension questions for students, which increased mental load but improved reading comprehension. Similarly, CoAsker \cite{liu2023student} encourages students to generate practice questions with an LLM-powered assistant, which were displayed in a side panel much like notes beside the document. When compared to generating practice questions without assistance, receiving questions that were generated by the assistant led to higher reading comprehension scores.

To our knowledge, no prior work has thoroughly explored integrating LLMs into the commenting feature of document reader software. The closest prior work are prototype mockups in Melin-Higgins' bachelor's thesis \cite{melin2024ai}, in which users select document text to issue pre-determined prompts to a hypothetical LLM-powered assistant. Responses appear as `sticky notes' beside the selected text, like margin notes, or they appear underneath the selected text by modifying the document structure. A very small 3-person study showed a preference for the margin note style. The study also only focused on usability and user preferences, and did not consider factors like reading comprehension. Furthermore, the prototypes were partially functioning Figma mockups, and none explored different levels of human and AI involvement.

\medbreak

Based on this research, AI margin notes could hinder or help readers. Creating them may limit deeper processing of the document text and lower comprehension. However, some AI margin note creation techniques may make note-taking less mentally demanding, so readers focus more on the underlying text while reading. Some techniques may even help readers take non-verbatim notes while improving motivation and cognitive engagement. 
Therefore, it is important to understand the effect of AI margin notes, and note-taking with LLMs more generally, on factors like test performance, workload, and user preferences. Our work contributes these important insights, which have been lacking in existing literature.

\begin{figure}[t]
    \centering
    \includegraphics[width=0.47\textwidth]{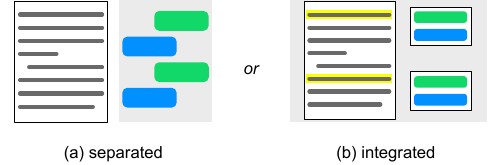}
    \caption{Integration: (a) a chat-based interface is not integrated since it is separated from the associated document text and (b) AI margin notes are integrated since they are associated to specific document text. \textit{In this and the following two figures: green denotes human-written text, like a prompt, and blue represents AI-generated text, like a response. Yellow indicates text that the AI margin note is linked to.}}
    \label{fig:integration}
\end{figure}

\section{AI Margin Notes}
We focus on three design parameters of AI margin notes: \emph{integration}, \emph{selection automation}, and the level of \emph{human and AI involvement}.

\subsection{Integration}
\label{sec:integration}
The primary difference between AI margin notes and chat-based tools is integration. Specifically, AI margin notes enhance \emph{comments} in document readers. Comments are anchored to specific text, making them integrated into the document, which contrasts with chat-based tools that are placed in a separate side panel with the content disconnected from the context (\autoref{fig:integration}). 
A separate interface for interacting with an AI assistant may suffice for general questions about the document or to receive overall summaries. However, readers also direct the AI assistant to specific parts of the text to contextualize their prompt \cite{kreijkes2025effects}. Separating these responses from the text can be inefficient.

First, \emph{referring to specific parts of a document while prompting requires additional work} \cite{DirectGPT}. Consider prompting an LLM to simplify a paragraph in a scientific document (e.g., \cite{GoogleGuidroz2025llm, AugustPaperPlain2023}). With a chat-based interface, the user must carefully formulate a prompt to refer to the specific paragraph (e.g., ``simplify the third paragraph in section 3''). Or, the user must select, copy, and paste the paragraph into the chat and write a prompt that uses \emph{deictic words} to refer to the paragraph instead (e.g., ``simplify \emph{this}''). With an AI margin note, the interaction is simpler as user can use deictic words without a separate copy and paste stage: they just select the paragraph and type ``simplify this'' where it appears in the document.

Second, \emph{shifting attention to a separate interface may distract from reading}. Note-taking requires frequent switching between reading and writing \cite{AdobeAcrobatAI, piolat2005cognitive}, and researchers suggest that note-taking should ``interrupt reading as little as possible'' \cite{marshall1997annotation}. Having readers frequently switch between separate reading and writing interfaces may further increase cognitive load \cite{harrison1995transparent}. These switching costs could be mitigated by presenting the prompting interface alongside the specific text the user is reading.

\begin{figure}[t]
    \centering
    \includegraphics[width=0.47\textwidth]{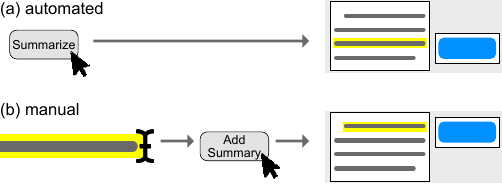}
    \caption{Selection automation: (a) text can be automatically selected and multiple AI margin notes can be created at once and (b) the user can manually select text to create an AI margin note.}
    \label{fig:automation}
\end{figure}

\begin{figure*}[t]
    \centering
    \includegraphics[width=\textwidth]{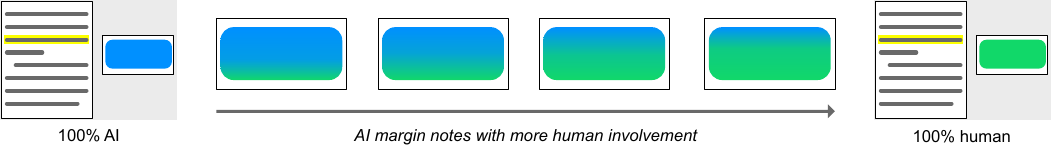}
    \caption{Levels of human and AI involvement: traditional margin notes consist of text that is 100\% written by a human, but an AI margin note could consist of text that is 100\% generated by AI, or could involve more human-generated text.}
    \label{fig:involvement}
\end{figure*}

Third, \emph{referring back to specific responses requires additional work}. For example, suppose a user prompted a chat-based LLM interface to get an explanation for an unfamiliar concept in a document. The explanation may be remembered in the short term, but if the user revisits the document much later, they must scroll through a lengthy and poorly organized chat history to find it, which may be especially difficult for readers with a lower working memory capacity to do \cite{ToScrollOrNot2009}. 
Linking responses directly to the relevant text avoids this issue. Some systems, like Adobe Acrobat Reader, support clicking on chat responses to highlight relevant parts of the document, however, this still requires scrolling through chat history.  Alternatively, the user could re-issue the prompt, but this wastes time and resources. Of course, users could copy responses they wish to save and paste them in comments \cite{kreijkes2025effects}, but this requires additional copy and paste steps that would not be needed with an AI margin note.

\subsection{Selection Automation}
AI margin notes are associated with specific text, which defines the context for the prompt and a location to display the note. Specifying text for the note could be done automatically by the AI assistant or manually by the user (\autoref{fig:automation}).
For example, consider summarization, a common task readers do in conventional margin notes \cite{jackson2001marginalia, porter2004beyond} and with LLM systems \cite{melin2024ai, kreijkes2025effects}. When using LLMs in document reader software, readers can generate a summary of the entire document automatically by pressing a single button \cite{AdobeAcrobatAI}, or they can manually specify which parts of the document need to be summarized by copying and pasting text into the chat. When the reader just presses a button, the LLM decides which information is important and worth including in the summary, but when the reader copies text to summarize, they are deciding which parts of the document are most important. Prior work on text highlighting suggests that this decision process could improve learning outcomes and reading comprehension, but it could also increase mental effort \cite{yue2015highlighting, joshi2024constrained}.

Automating the selection of text to create AI margin notes may also impact psychological ownership, feelings of the learning experiences \emph{belonging} to them \cite{Pierce2001PsychologicalOwnership}. Prior work shows that fostering psychological ownership, which can be achieved by giving learners more \emph{control} over their learning \cite{Pierce2001PsychologicalOwnership, Pierce2003PsychologicalOwnership}, can encourage more active learning \cite{griffith1990cooperative} and motivates learning  \cite{soslau2007urban}. Automatically generating AI margin notes reduces effort, but potentially at the risk of lowering comprehension and psychological ownership.

\subsection{Human and AI Involvement}
\label{sec:humanAIInvolvement}
Currently, commenting in document reader software is intended for text that is entirely written by a human. Specifically, the user must formulate their own ideas and type into a text box. At the other extreme, pressing a ``Summarize'' button in a document reader \cite{AdobeAcrobatAI} produces text that is entirely written by the LLM without any guidance from the human. However, there are techniques that require involvement from both the human and the LLM. For example, a chat-based interface requires the user to provide a specific instruction to the LLM, such as ``summarize the text for someone in high school.'' 
Here, the user \rev{forms goals and sub-tasks \cite{MetacognitiveDemandsPrompting2024} to exert some control over the output \cite{ConstrainedGPT}, and the LLM produces it}. These roles can be reversed too, for example, the LLM could `prompt' the user to respond to practice reading comprehension questions \cite{wang2024chatprcs, liu2023student}.

AI margin notes can also vary in how much human and AI involvement they require to produce the final comment text (\autoref{fig:involvement}). 
Techniques that require more human involvement likely require more cognitive engagement than those with more AI involvement. This may be beneficial for reading comprehension \cite{bjork1994memory, bjork2011making, bjork1994institutional, tyler1979cognitive}, may discourage readers from taking verbatim notes \cite{bretzing1979processing}, and may even improve feelings of psychological ownership \cite{ConstrainedGPT, joshi2025interactiontechniquesencouragelonger}.
However, too much cognitive engagement can frustrate and discourage learners \cite{grund2024learning}, which may be mitigated by increasing AI involvement.

\begin{figure*}
    \centering
    \includegraphics{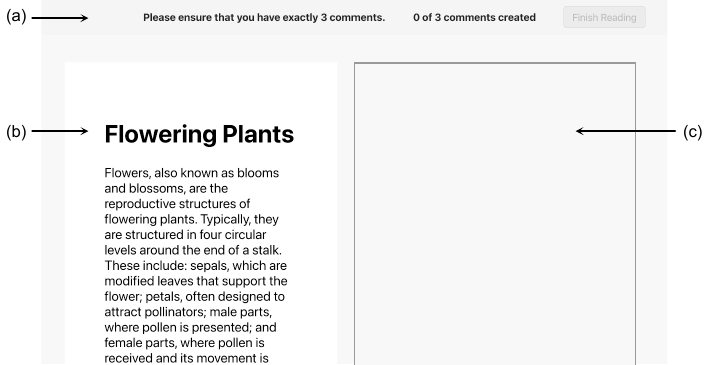}
    \caption{Experimental reading interface: (a) toolbar containing instructions, (b) document to read, and (c) space for specific design variations to be displayed.}
    \label{fig:interface}
\end{figure*}

\section{Experimental Method}
We conducted \rev{three experiments to explore these three design parameters of AI margin notes}. Specifically, we were interested in understanding how integration, selection automation, and human and AI involvement affect reading comprehension, duration, psychological ownership, task workload, and user preferences. All experiments \rev{used the same experimental method and} were conducted using the Prolific crowdsourcing platform,\footnote{\href{https://www.prolific.com}{https://www.prolific.com}} \rev{but with different participants for each experiment}. \rev{Note our protocol was reviewed and approved by our institution's Research Ethics Board.}

\subsection{Task}
The primary experimental task was composed of two stages. First, a \emph{reading stage}, where participants read non-fiction documents that were approximately 500 words each. For each document, they interacted with an LLM assistant using a different technique that represents a variation within the design parameter under evaluation. Second, a \emph{test stage}, where participants completed reading comprehension tests about each document. Each test consisted of six multiple choice questions. The documents and questions were developed by Wallace et al. \cite{WallaceReadingPassages} and were designed by a learning and reading specialist to be suitable for an eighth grade reading level.\footnote{Note that we slightly modified the wording of some questions, as we noticed that some questions could be successfully answered by looking at the wording of other questions. Our versions of these questions and the associated documents are included in the supplementary materials. At the time of writing this paper, Wallace et al.'s license agreement permits reuse, modification, public display, and redistribution for non-commercial research purposes.} This is representative of documents targeted for the general public \cite{nngroupLegibilityReadability}.

\subsection{Apparatus}
Our experimental system was a custom Node.js and React web application that implemented two types of interfaces, one for each stage: the reading interface, where participants read a document and interacted with an LLM using a specific technique (\autoref{fig:interface}); and the testing interface, where participants answered time-bounded, multiple-choice reading comprehension tests. \emph{The supplementary video demonstrates both interfaces.}

\subsubsection{Reading Interface}
The document was displayed on the left side of the screen. The specific design variation of the experimental condition was displayed beside it (details provided in each individual experiment). All design variations that involved an LLM used GPT 4.1 mini.\footnote{The system prompts used to generate responses are included in the supplementary materials.}
The top of the screen contained a toolbar indicating how many comments or responses they still had to complete and a blue ``Finish Reading'' button to end the trial.

\subsubsection{Testing Interface}
The testing interface displayed six multiple choice questions at the centre of the screen. At the top of the screen was a toolbar that displayed how many questions the participant had answered, a countdown timer (displayed numerically and as a progress bar that shrank every second), and a blue ``Finish Test'' button that could be pressed to end the test.

\subsection{Procedure}
Participants received a link to an experimental web application through Prolific. 
They had to use a desktop or laptop computer, which was strictly enforced through the web application. \rev{First, participants read a consent form, which detailed inclusion and exclusion criteria, data handling procedures (anonymized and stored on encrypted hard drives), and remuneration. After providing informed consent, participants entered basic demographic information and read instructions.\footnote{\rev{The demographics questionnaire and all instructions are included in the supplementary materials.}}} \rev{The instructions described the general nature of the reading and testing stages, and details on how to use the specific techniques being tested.}

Next, they completed the reading stage. There was no time limit, but to focus on the effect of the different techniques rather than the number of responses, participants had to leave 3 AI margin notes (or receive 3 responses from the AI assistant, depending on the technique). 
After reading the document, they answered 10 questions about their experience and 1 question about their prior knowledge of the document's content. They repeated this for the other documents, then answered 2 questions about their overall preferences.

Two hours later, the participant was invited to return for the test stage. Given the lower reading difficulty of the documents, we increased the difficulty of the test by making it closed-book and restricted to 60 seconds. After completing all tests, participants described other study aids they used, such as taking a screenshot of the document or writing notes outside of the reading interface.

\begin{table*}[t]
    \centering
    \caption{\rev{ Experiment 1 demographics.}}
    \small %
\begin{tabular}{lr|lr|lr}
\toprule
\multicolumn{2}{l|}{Gender} & \multicolumn{2}{l|}{Age} & \multicolumn{2}{l}{Education} \\
\midrule
Men & 16 & 25-34 & 4 & High School & 2\\
Women & 10 & 35-44 & 14 & Some University (no credit) & 5\\
&& 45-54 & 3 & Bachelor's Degree & 14 \\
&& 55-64 & 3 & Professional Degree Beyond Bachelor's & 1 \\
&& 65-74 & 2 & Master's Degree  & 3\\
&&&& Doctorate Degree & 1 \\
\bottomrule
\end{tabular}

\begin{tabular}{lr|lr|lr|lr}
\\
\toprule
\multicolumn{2}{l|}{Document Reader Frequency} & \multicolumn{2}{l|}{Commenting Frequency} &\multicolumn{2}{l|}{LLM Frequency} & \multicolumn{2}{l}{LLM Summarization Frequency}\\
\midrule
Daily & 6 & Daily & 1 & Daily & 10 & Daily & 3\\
Weekly & 11 & Weekly & 6 & Weekly & 9 & Weekly & 9\\
Monthly & 7 & Monthly & 3 & Monthly & 5 & Monthly & 3\\
Less than Monthly & 2 & Less than Monthly & 5 & Less than Monthly & 5 & Less than Monthly & 5\\
&& Never & 11 & Never & 1 & Never & 6\\

\bottomrule

\end{tabular}
    \label{tab:exp1Demographics}
\end{table*}

\subsection{Design}
All experiments use within-subjects design since the documents were relatively short, and we wanted participants to compare their experiences with each technique. There is one independent variable, \f{condition}, which was randomly assigned. For a single \f{condition}, one of six documents from Wallace et al. \cite{WallaceReadingPassages} was randomly-assigned. For increased internal validity, a single document was not restricted to a single condition and could be assigned to any condition. 

\subsection{Measures}
Reading comprehension and duration were calculated from logs, and all other metrics were calculated from subjective questionnaires completed after the reading stages. \rev{With the exception of rankings,} these used a 0-100 interval range.\footnote{The supplementary materials contains all questions.}

\subsubsection{Reading Comprehension}
This represents the number of questions that were correctly answered during the test stage (0-6 range). Analyzing \m{Reading Comprehension} gives insights into the effects of AI margin notes on overall learning, and whether techniques that encourage more human involvement improve learning outcomes.

\subsubsection{Duration}
This is the time taken in minutes to complete the reading stage. Some AI margin notes that require more manual effort or human involvement likely take longer to complete.

\subsubsection{Psychological Ownership}
This was measured by asking two quantitative questions about \m{Personal Ownership} and \m{Responsibility} \cite{Caspi2011Collaboration}. The average of the two create a composite measure, as done in prior work \cite{joshi2025interactiontechniquesencouragelonger, ConstrainedGPT}. 
The internal consistency reliability score, which describes how consistently different items on a questionnaire describe the same underlying concept, was high for all experiments (.80 $\leq \alpha \leq$ .90), suggesting that the composite measure was appropriate to use for data analysis. As described previously, this is an important measure since fostering psychological ownership can improve learner experiences.

\subsubsection{Task Workload}
We used factors from the NASA-TLX \cite{NASATLX}: \m{Mental Demand}, \m{Physical Demand}, \m{Temporal Demand}, \m{Performance}, \m{Effort}, and \m{Frustration}. 
Techniques that are more manual or that require more human involvement may involve more mental demand and effort, which could be beneficial for learning \cite{bjork1994memory, bjork2011making, bjork1994institutional, tyler1979cognitive}, however, this may not happen if they are too frustrating \cite{grund2024learning}.

\subsubsection{Preferences}
We asked about \m{Frequency of Use}, a question from the system usability scale (SUS) \cite{brooke1996sus} representing how frequently the participant would use the feature if made available to them.
After trying all techniques, participants also gave each \f{condition} a \m{Ranking}, where 1 was the best, 2 was the second best, and so on. Ties were allowed. To establish an \m{Overall Ranking} of conditions, we use the Condorcet voting method \cite{young1988condorcet}.
An overall rank of 1 means that technique `defeats' all others in pairwise comparisons. An overall rank of 2 means that technique `defeats' all others except for the technique ranked first, and so on.
This is important to study, as users would likely want to keep using techniques that they like, which is especially useful if it also improves their learning outcomes.

\rev{
\subsubsection{Other Metrics}
We triangulate these measures with other data, such as characteristics of participant prompts and the length and location of selected text. These metrics are specific to each experiment, so we introduce them with their results.
}

\section{Experiment 1: Integration}
The goal of this experiment is to understand the effect of integration that is achieved with AI margin notes. Participants wrote prompts using a traditional, chat-based interface (\f{chat}), or an AI margin note (\f{note}).

\begin{figure*}[t]
    \centering
    \includegraphics{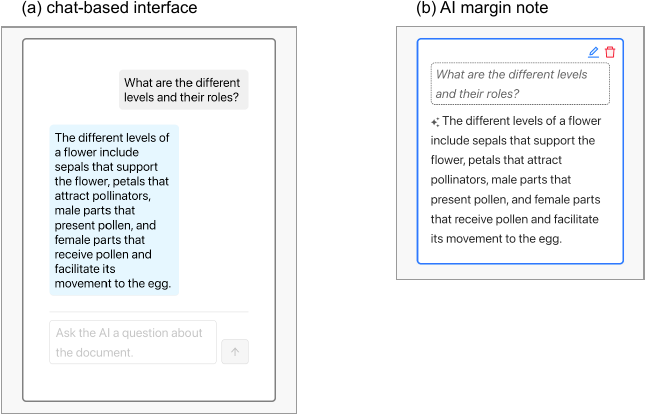}
    \caption{Techniques tested in Experiment 1: (a) a chat-based interface and (b) AI margin notes.}
    \label{fig:exp1Techs}
\end{figure*}

\subsection{Participants}
We recruited 29 participants through Prolific. Participants were restricted to the United States and Canada, and those who had completed 2,500 previous tasks on the platform with a 99-100\% approval rating. Three participants (10\%) were removed for using other study aids prior to the test, leaving 26 valid responses (\autoref{tab:exp1Demographics}). All self-reported proficiency in reading in English. Each participant received \$10, plus a \$5 bonus for scoring in the top 25\% on the comprehension test, to encourage more conscientious active reading. 
The experiment took roughly 25 minutes in total.

\subsection{Apparatus} 
The reading interface had two variations. For the chat-based prompting technique, a chat interface was displayed to the right of the document. Here, participants could type prompts in a text box and press a blue ``Submit'' button. Their prompts appeared like a chat message on the right in a grey bubble, and LLM responses were displayed on the left in a light blue bubble (\autoref{fig:exp1Techs}a).

To make more direct comparisons between the two variations, we had to create an AI margin note technique where participants type prompts and receive responses from an LLM.
To create an AI margin note, the participant selected text in the document using their cursor, which caused a blue ``Comment'' button to appear. Clicking this created a text box to appear that was anchored to part of the document text and resembled a Google Docs comment (\autoref{fig:exp1Techs}b). The participant could type a prompt and press a blue ``Confirm'' button. After a few seconds, the LLM provided a response. The participant's prompt appeared at the top of the comment in grey, italicized text with a dashed border around it, and the response appeared below in black text, prepended with a black `sparkle' icon. This styling clearly distinguishes human-written text (no icon) from AI-written text (with a `sparkle' icon) and main comment text (black text) from additional context that helped generate the comment (grey, italicized text with a dashed border). Participants could edit their prompt to receive new output and they could delete their comments. Clicking on individual comments highlighted the corresponding text in the document.

\subsection{Results}
We use Wilcoxon signed-rank tests to investigate the effects of \f{condition} on the various measures. To streamline the presentation of results, \emph{details of these statistical tests are shown in \autoref{tab:exp1Stats}} and \emph{all data is included in the supplementary materials}.

Overall, there were no significant effects of \f{condition} on \m{Reading Comprehension}, \m{Duration}, \m{Psychological Ownership}, any of the workload-related factors, and \m{Frequency of Use}. 

\subsubsection{Preferences}
Although there were no differences for the aforementioned measures, most participants preferred prompting with AI margin notes. From the Condorcet voting method, we observed that \f{note} received an \m{Overall Ranking} of 1, and \f{chat} received an \m{Overall Ranking} of 2. Notably, the majority of participants (17, 65\%) assigned \f{note} a \m{Ranking} of 1, and \f{chat} a \m{Ranking} of 2 (16, 62\%).

To better understand why participants preferred \f{note} over \f{chat}, we examined participants' free-form responses. Several (12, 46\%) mentioned how \f{note} was easier and more intuitive than \f{chat}. They described how they \ppquote{liked [not] having to move to a separate chat}{P21}, how \ppquote{having the comment embedded right into the document [was] handy and easy to reference along with the actual text}{P5}, and how \ppquote{selecting text [was] more natural when one has a question related to that part of the text}{P14}.

\rev{
\subsubsection{Prompt Wording}
These responses regarding the ease of referring to specific parts of the document were corroborated by examining how participants worded their prompts. Participants did not have to write prompts that were as specific when they used \f{note}, perhaps due to the increased specificity enabled by text selections. Notably, 29 prompts (37\%) written with \f{note} referred to specific nouns in the document using deictic words (e.g., ``this,'' ``they,'' ``it,'' and ``here''). With \f{chat}, deictic words were less frequent (17, 22\%) and most instances referred to specific nouns from previously-entered prompts (10, 13\%). 
For example, consider P3 and P21, who both asked follow-up questions about materials used to make wagon wheels. P21, who read this document with \f{chat}, asked \ppquote{what were wagon wheels made of before iron?} But P3, who used \f{note}, instead selected text that discussed wagon wheel material, and asked \ppquote{how much of an impact did \textbf{this} have for the world?}
}

\subsection{Summary}
Overall, our results suggested that the integration of AI margin notes is highly desirable and preferable over chat-based interfaces as it prevented switching between interfaces; it was easier to reference individual comments later; \rev{and it was easier to to ask questions using deictic words to refer to specific nouns described in the selected text.}
In this experiment, associating text to the AI margin note was done \emph{manually} by the participant. However, selecting text could also be done \emph{automatically} by the LLM, which we explore further in the following experiment.

\begin{table*}[t]
    \centering
    \caption{\rev{Experiment 2 demographics.}}
    \small %
\begin{tabular}{lr|lr|lr}
\toprule
\multicolumn{2}{l|}{Gender} & \multicolumn{2}{l|}{Age} & \multicolumn{2}{l}{Education} \\
\midrule
Men & 18 & 25-34 & 4 & High School & 4\\
Women & 12 & 35-44 & 15 & Some University (no credit) & 7\\
&& 45-54 & 8 & Bachelor's Degree & 9 \\
&& 55-64 & 1 & Master's Degree & 10 \\
&& 65-74 & 1 \\
&&75+&1&  \\
\bottomrule
\end{tabular}

\begin{tabular}{lr|lr|lr|lr}
\\
\toprule
\multicolumn{2}{l|}{Document Reader Frequency} & \multicolumn{2}{l|}{Commenting Frequency} &\multicolumn{2}{l|}{LLM Frequency} & \multicolumn{2}{l}{LLM Summarization Frequency}\\
\midrule
Daily & 4 & Daily & 1 & Daily & 8 & Daily & 2\\
Weekly & 10 & Weekly & 3 & Weekly & 12 & Weekly & 8\\
Monthly & 7 & Monthly & 4 & Monthly & 5 & Monthly & 4\\
Less than Monthly & 7 & Less than Monthly & 3 & Less than Monthly & 5 & Less than Monthly & 11\\
Never & 2 & Never & 18 & Never & 2 & Never & 5\\

\bottomrule

\end{tabular}
    \label{tab:exp2Demographics}
\end{table*}

\section{Experiment 2: Selection Automation}
\rev{After learning that the integration of AI margin notes through text selections was desirable, we then wanted to learn \emph{how} such margin notes should be associated to specific text selections.}
The goal of this experiment is to understand the effect of selection automation when creating AI margin notes. Participants either manually selected what text should be summarized (\f{manual}), or pressed a button to let the AI assistant place three summary AI margin notes in the document (\f{automatic}).

\subsection{Participants}
Using the same inclusion criteria as Experiment 1, we recruited 32 new participants on Prolific. Two participants (6\%) were excluded for not attempting to answer any reading comprehension questions, or for using other study aids, leaving 30 valid responses (\autoref{tab:exp2Demographics}). All self-reported English reading proficiency. \rev{As in Experiment 1, participants received \$10 with a \$5 bonus incentive for top-25\% comprehension test performance}. The experiment took approximately 25 minutes in total.

\subsection{Apparatus}
\begin{figure*}[t]
    \centering
    \includegraphics{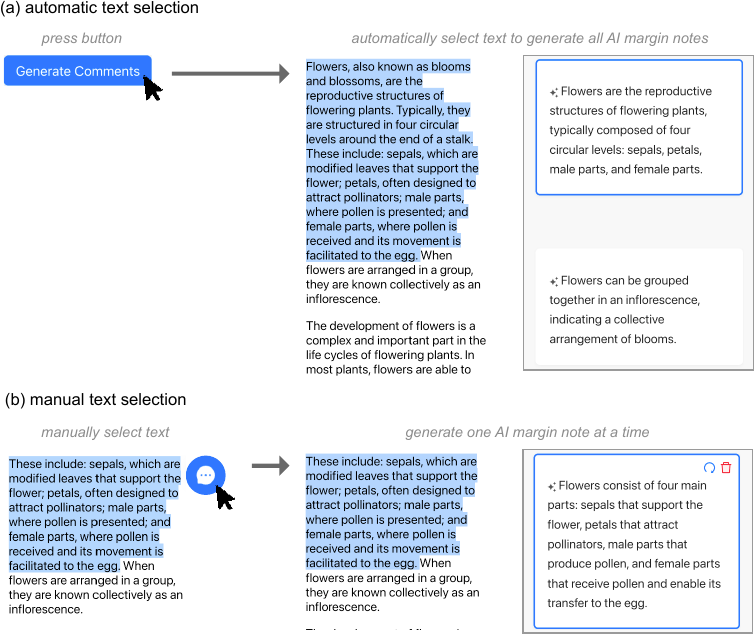}
    \caption{Techniques tested in Experiment 2: (a) automatically selecting text to generate all AI margin notes at once and (b) manually selecting text and generating AI margin notes one-by-one.}
    \label{fig:exp2Techs}
\end{figure*}

Both variations focused specifically on producing AI margin notes that summarized text as this is a representative task that is currently done manually, by the user copying and pasting specific parts of a document into a chat-based interface, or automatically, by pressing a button. 
For the automatic selection technique, the top toolbar displayed a blue ``Generate Comments'' button in the top left corner (\autoref{fig:exp2Techs}a). Pressing this caused three AI margin note comments that were linked to the document text to appear to the right of the document. Each summarized specific selections from the document, and was displayed in black text prepended with a black `sparkle' icon. As before, participants could click individual comments to highlight the corresponding text, but they could not delete or edit individual comments. Instead, they could press the ``Generate Comment'' button again to regenerate all comments. 

The manual selection technique worked like the previous experiment (\autoref{fig:exp2Techs}b): the participant selected text from the document and pressed a blue ``Comment'' button to generate a summary comment (appearing after a few seconds). The styling of the comment was the same, except that participants could also delete or regenerate individual comments. The participant had to manually create three summary comments.

\begin{figure*}
    \centering
    \includegraphics[width=\textwidth]{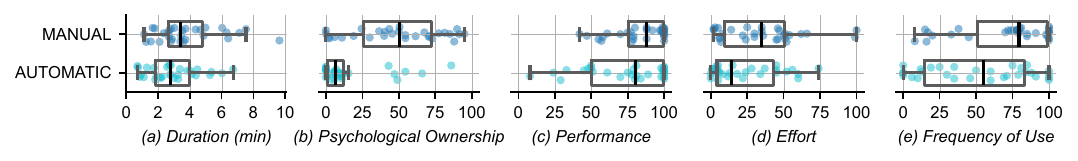}
    \caption{Experiment 2 results: (a) \m{Duration}, (b) \m{Psychological Ownership}, (c) \m{Performance}, (d) \m{Effort}, and (e) \m{Frequency of Use}.}
    \label{fig:exp2Results}
\end{figure*}

\begin{figure}
    \centering
    \includegraphics[width=0.47\textwidth]{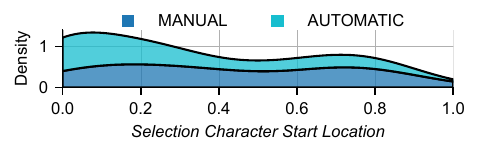}
    \caption{\rev{Distribution of selection start locations in normalized document position, shown using a kernel density estimate, where Density indicates the estimated concentration of points.}}
    \label{fig:selectionPosition}
\end{figure}

\subsection{Results}
As before, we use Wilcoxon signed-rank tests where applicable. \emph{Details of statistical tests are shown in \autoref{tab:exp2Stats}}. 

We did not observe any significant effect of \f{condition} on \m{Reading Comprehension}, so we focus our results on other metrics.

\subsubsection{Duration}
Participants were 39 seconds slower when they manually selected text (\autoref{fig:exp2Results}a).
There was a significant effect of \f{condition} on \m{Duration}, revealing that \f{manual} (\median{3.43}, \iqr{2.15}) was slower than \f{automatic} (\median{2.78}, \iqr{2.14}).

\subsubsection{Psychological Ownership}
Though slower, participants generally felt more psychological ownership when they manually selected text (\autoref{fig:exp2Results}b). A significant effect of \f{condition} on \m{Psychological Ownership} revealed that participants felt more \m{Psychological Ownership} with \f{manual} (\median{50.25}, \iqr{46.62}) than \f{automatic} (\median{6.75}, \iqr{10.88}). This was supported by free-form responses like \ppquote{[manually selecting text] requires me to focus more on the task, [which] gives me more responsibility}{P20}.

\subsubsection{Task Workload}
Participants felt like they performed better when selecting text manually, despite it requiring more effort. Specifically, there was a significant effect of \f{condition} on \m{Performance} (\autoref{fig:exp2Results}c), suggesting that participants felt like they were better at creating AI margin notes with \f{manual} (\median{88}, \iqr{24.25}) than \f{automatic} (\median{80.5}, \iqr{49.5}). 

A significant effect of \f{condition} on \m{Effort} (\autoref{fig:exp2Results}d) suggested that participants felt more \m{Effort} for \f{manual} (\median{35}, \iqr{41.25}) than \f{automatic} (\median{14}, \iqr{38.75}). This was supported by free-form responses like \ppquote{in being able to highlight on my own, I also feel as if I did some of the work and not just nothing}{P25}. We did not observe any differences between \f{manual} and \f{automatic} for \m{Mental Demand}, \m{Physical Demand}, \m{Temporal Demand}, and \m{Frustration}.

\subsubsection{Preferences}
Participants generally seemed to prefer generating AI margin notes manually. There was a significant effect of \f{condition} on \m{Frequency of Use} (\autoref{fig:exp2Results}e), suggesting a preference to use \f{manual} (\median{79.5}, \iqr{47.5}) more frequently than \f{automatic} (\median{55}, \iqr{68.75}). For \m{Overall Ranking}, we observed that participants tended to prefer \f{manual}, which was ranked first. Specifically, 23 participants (77\%) gave \f{manual} a \m{Ranking} of 1, and 21 (70\%) gave \f{automatic} a \m{Ranking} of 2.

A few (4, 13\%) mentioned how \f{manual} provided more incentive to read the document, for example: \ppquote{generating all summaries at once will encourage users not to actually read the document for themselves. Generating individual summaries after I selected text meant I [had] to actually read the document}{P10}. The majority (16, 53\%) appreciated \f{manual} for the increased control it enabled, for example: \ppquote{I felt a little more control [when] I was able to pick what I wanted to be summarized. [When] I just pressed one button, it summarized everything, but not necessarily the areas I wanted it to}{P22}.

\begin{figure*}[t]
    \centering
    \includegraphics{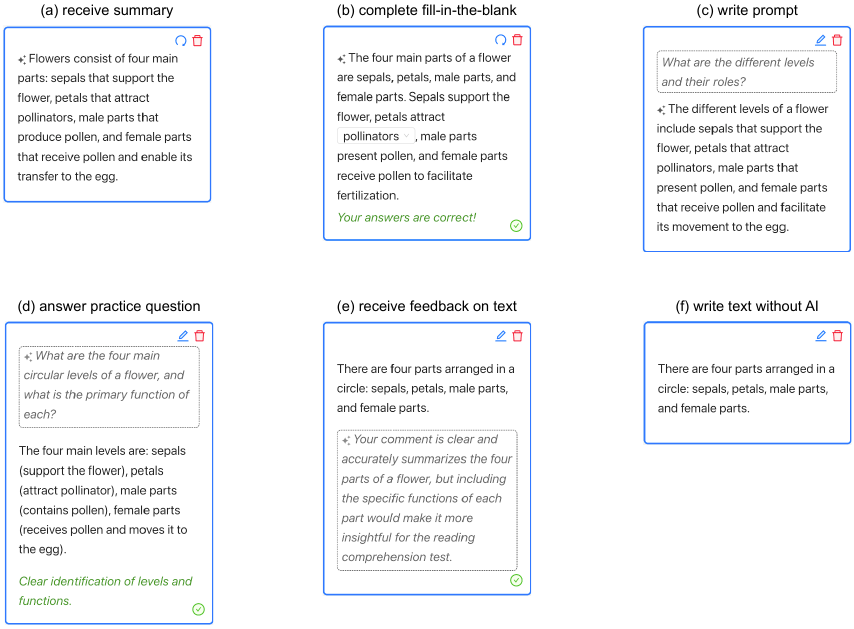}
    \caption{Techniques tested in Experiment 3: (a) receive a summary (100\% AI-written text), (b) complete a fill-in-the-blank exercise, (c) write a prompt, (d) answer a practice question, (e) receive feedback on written text, and (f) write a comment without AI (100\% human-written text).}
    \label{fig:continuum}
\end{figure*}

\subsubsection{Selection Location and Length}
\rev{
The location and length of selections supports participant comments that \f{manual} encouraged reading the document and enabled more control over selected text. 
We examined the distribution of \textit{Selection Location} (\autoref{fig:selectionPosition}) and found the LLM tended to place comments at the beginning of the document with \f{automatic}, while participants placed them more consistently throughout the document with \f{manual}.
We also calculated the \m{Selection Word Count}, finding selections were significantly shorter with \f{manual} (\median{62}, \iqr{83}) compared to \f{automatic} (\median{83}, \iqr{90.75}). 
}

\subsection{Summary}
Overall, our results suggested that selecting text for an AI margin note is slower and requires more effort when done manually. However, this may be a worthwhile trade-off, as manually selecting text was associated with higher feelings of psychological ownership, better perceived performance, and was preferred by participants. \rev{This may have been due in part to the increased control over where text selections were placed, as automatic text selections tended to be placed at the beginning of the document}.
In this experiment, the resulting AI margin note was always a summary, however, there are many other types of AI margin notes that can be created that involve the human and AI in different ways.

\section{AI Margin Note Techniques}
To fully explore the design parameter of human and AI involvement, we implemented six techniques that fall along a continuum, ranging from AI margin notes whose text was generated entirely by the LLM, to those whose text was written entirely by the user (\autoref{fig:continuum}). All are created by selecting text manually and pressing a blue ``Comment'' button, and \emph{all are demonstrated in the supplementary video.}

\begin{table*}[t]
    \centering
    \caption{\rev{Experiment 3 demographics.}}
    \small %
\begin{tabular}{lr|lr|lr}
\toprule
\multicolumn{2}{l|}{Gender} & \multicolumn{2}{l|}{Age} & \multicolumn{2}{l}{Education} \\
\midrule
Men & 13 & 18-24 & 1 & High School & 7 \\
Women & 19 & 25-34 & 2 & Some University (no credit) & 2 \\
&& 35-44 & 7 & Technical Degree & 5\\
&& 45-54 & 13 & Bachelor's Degree & 14 \\
&& 55-64 & 4 & Master's Degree & 4 \\
&& 65-74 & 4 \\
&& 75+ &1&  \\
\bottomrule
\end{tabular}

\begin{tabular}{lr|lr|lr|lr}
\\
\toprule
\multicolumn{2}{l|}{Document Reader Frequency} & \multicolumn{2}{l|}{Commenting Frequency} &\multicolumn{2}{l|}{LLM Frequency} & \multicolumn{2}{l}{LLM Summarization Frequency}\\
\midrule
Daily & 6 & Daily & 1 & Daily & 13 & Daily & 3\\
Weekly & 14 & Weekly & 6 & Weekly & 11 & Weekly & 12\\
Monthly & 8 & Monthly & 4 & Monthly & 5 & Monthly & 3\\
Less than Monthly & 3 & Less than Monthly & 8 & Less than Monthly & 3 & Less than Monthly & 10\\
Never & 1 & Never & 12 & &  & Never & 4\\

\bottomrule

\end{tabular}
    \label{tab:exp3Demographics}
\end{table*}

\subsection{Receive Summary}
The technique with the least human involvement is receiving a generated summary where all of the text is written by AI (\autoref{fig:continuum}a). This \rev{represents a common way LLMs are used in document readers and} is identical to the manual text selection technique presented in Experiment 2.

\subsection{Complete Fill-in-the-Blank Exercise}
A technique with slightly more human involvement is receiving a generated summary, but with some keywords omitted to resemble a fill-in-the-blank question (\autoref{fig:continuum}b). Like the previous technique, the LLM produces a summary for the text selected by the user, shown in black text and prepended by a black `sparkle' icon. Each summary contains one or two `blanks,' displayed using dropdowns containing three options. The user must select the correct answer from the provided options. Underneath, the comment displays feedback to notify the user whether their responses are correct (shown in green with a check mark) or incorrect (shown in red with an `X'). The user can delete the comment or regenerate it. \rev{Fill-in-the-blank questions are a common way of assessing reading comprehension \cite{WilsonCloze1953, nngroupClozeTest} as they require readers to infer the contents of missing words based on their own understanding of the text \cite{jonz1994effects}.}

\subsection{Write Prompt}
A technique with some human and some AI involvement is writing a custom prompt for the LLM to respond to (\autoref{fig:continuum}c). 
\rev{Writing prompts is the primary way users interact with current chat-based tools and allows the user to achieve goals that go beyond summarization, such as learning additional information about the text or simplifying it. Formulating a prompt also requires users to form goals and subtasks, which may encourage them to reflect on their own understanding of the text \cite{MetacognitiveDemandsPrompting2024}.}
This is identical to the AI margin note technique described in Experiment 1.

\subsection{Answer Practice Question}
Another technique with some involvement from both the human and the AI is answering a practice reading comprehension question that was generated by the LLM, which has been shown to help with learning \cite{wang2024chatprcs, liu2023student}. The LLM produces a practice short-answer question based on the selected text (\autoref{fig:continuum}d), prepended with a grey `sparkle' icon and shown using grey italicized text with a dashed border. The user types in their answer in a text box underneath and presses a blue ``Confirm'' button to submit their response. The submitted response is then shown in black text. After a few seconds, the user receives feedback from the LLM about their answer: correct responses are shown in green and with a check mark underneath the response, and incorrect responses are shown in red with an `X.' All responses have to be correct. The user can edit their responses to correct mistakes (and will receive new feedback accordingly), or can delete their comment.

\subsection{Receive Feedback on Written Text}
A technique that has a lot of human involvement is requiring users to write their own comment text, with LLM-generated feedback and suggestions on how their comment can be improved (\autoref{fig:continuum}e). This can be beneficial, as many learners struggle to take effective notes \cite{bretzing1979processing}. \rev{Prior work suggests that when writing, AI-generated feedback and suggestions can help users see alternative perspectives and encourage reflection \cite{DangBeyondTextGeneration2022, WriterPersonas2024}, which could be helpful in the context of note-taking}. The user can type within the provided text box and press ``Confirm.'' The comment text is shown as black text. After a few seconds, the AI assistant provides feedback on the text, prepended with a grey `sparkle' icon and shown as grey, italicized text with a dashed border. A green check mark is shown if their comment text is `good,' and a red `X' is shown if it is not. All comment text has to be classified as good. The user can edit their comment to receive updated feedback and can delete it.

\subsection{Write Text without AI}
The technique with the most human involvement is requiring the user to write their own comment, without any AI assistance (\autoref{fig:continuum}f). This baseline resembles the behaviour of existing commenting systems, where all of the text is written by the user. The user can type within the provided text box and press ``Confirm.'' The comment is shown as black text. The user can edit or delete the comment.

\begin{figure*}[t]
    \centering
    \includegraphics[width=\textwidth]{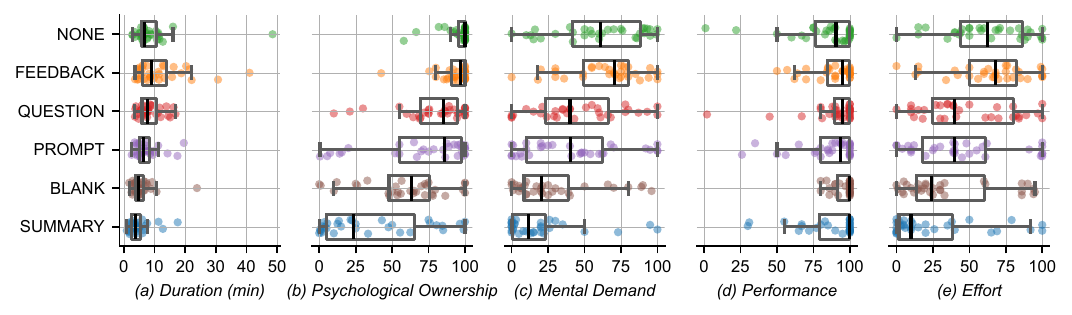}
    \caption{Experiment 3 results: (a) \m{Duration}, (b) \m{Psychological Ownership}, (c) \m{Mental Demand}, (d) \m{{Performance}}, and (e) \m{Effort}.}
    \label{fig:exp3Results}
\end{figure*}

\section{Experiment 3: Human and AI Involvement}
\rev{The previous experiment found that manual text selection is preferred, but it did not explore different ways to generate AI margin notes after selection.}
The goal of this experiment is to investigate the effect of human and AI involvement when generating the AI margin note itself. Participants read six documents, one for each technique described above: receiving a summary (\f{summary}), completing a fill-in-the-blank exercise (\f{blank}), writing a prompt (\f{prompt}), answering a practice question (\f{question}), receiving feedback on written text (\f{feedback}), and writing without any AI assistance (\f{none}). The experiment was longer, approximately 90 minutes.

\subsection{Participants}
We recruited 36 new participants on Prolific, using the same inclusion criteria as the previous experiments. Four (11\%) were excluded for experiencing technical issues, leaving 32 valid responses (\autoref{tab:exp3Demographics}). All but one were proficient at reading in English. Participants received \$25 total, with an additional \$5 bonus if their total reading comprehension score was within the top 25\%.

\subsection{Results}
Where applicable, we use Friedman omnibus tests and Wilcoxon signed-rank post hoc tests, with Holm's corrections for multiple comparisons; and Spearman's correlations. \emph{Statistical test details are shown in \autoref{tab:exp3Stats}}.

\subsubsection{Reading Comprehension}
Contrary to what prior work suggests, techniques with more human involvement were not associated with higher reading comprehension scores. Specifically, the differences between \f{summary} (\median{4}, \iqr{2.25}) and all other conditions (all \median{5}, \iqr{2}) are marginal (\p{.054}{=}), and not statistically significant.

\subsubsection{Duration}
Participants were faster when they used techniques that did not require as much human involvement (\autoref{fig:exp3Results}a). A significant effect of \f{condition} on \m{Duration} revealed that \f{summary} (\median{3.96}, \iqr{2.78}) and \f{blank} (\median{4.92}, \iqr{2.80}) were faster than all other techniques.

\subsubsection{Psychological Ownership}
Generally, participants felt more psychological ownership for techniques that required more human involvement (\autoref{fig:exp3Results}b). A significant effect of \f{condition} on \m{Psychological Ownership} and post hoc tests revealed that \f{none} (\median{100}, \iqr{4.38}) and \f{feedback} (\median{97}, \iqr{9.34}) were associated with the highest \m{Psychological Ownership}, and \f{summary} (\median{23.75}, \iqr{60.88}) with the lowest.

\subsubsection{Task Workload}
Generally, techniques that required more human involvement were more mentally demanding, associated with poorer perceived performance, and more effortful. For \m{Mental Demand}, there was a significant effect of \f{condition} (\autoref{fig:exp3Results}c) and post hoc tests showed that \f{none} (\median{61}, \iqr{47}) and \f{feedback} (\median{70.5}, \iqr{31}) were associated with higher scores than \f{prompt} (\median{40.5}, \iqr{52.25}), \f{blank} (\median{20.5}, \iqr{30.75}), and \f{summary} (\median{11.5}, \iqr{22.25}). 
There was also a significant effect of \f{condition} on \m{Performance} (\autoref{fig:exp3Results}d), with post hoc tests revealing that participants generally felt like they did better at the task with \f{blank} (\median{100}, \iqr{8.5}) than \f{none} (\median{90.5}, \iqr{23.5}), \f{feedback} (\median{95}, \iqr{15.25}), and \f{summary} (\median{100}, \iqr{20.75}).
For \m{Effort}, a significant effect of \f{condition} (\autoref{fig:exp3Results}e) and post hoc tests showed that \f{none} (\median{62.5}, \iqr{42.75}) and \f{feedback} (\median{68}, \iqr{32.5}) had the highest scores, and \f{summary} (\median{10}, \iqr{36.75}) had the lowest.
We did not observe significant differences between the different techniques for \m{Physical Demand}, \m{Temporal Demand}, and \m{Frustration}.

\subsubsection{Preferences}
There were no differences in \m{Frequency of Use}, with all medians being 50 or greater, suggesting that participants would like to use all techniques. For \m{Overall Ranking}, the Condorcet voting method revealed that overall, \f{blank} was ranked first, followed by \f{question}, \f{summary}, \f{prompt}, \f{feedback}, and \f{none}. This \m{Overall Ranking} suggests that participants generally preferred techniques with more AI involvement (\f{blank}, \f{question}, \f{summary}, and \f{prompt}) over those with more human involvement (\f{feedback} and \f{none}), for example: \ppquote{the more involved AI was, the more I appreciated the help}{P15}.

Considering the \m{Ranking} scores, 21 (66\%) ranked \f{blank} within the top 3, and 20 (62\%) placed \f{question} within the top 3. These two techniques were generally valued as they \ppquote{had the greatest pedagogical heft}{P7} and \ppquote{actively engage memory and reinforce understanding}{P4}. Nineteen (59\%) placed \f{summary} within the top 3, however, eight (25\%) gave it a rank of 6, suggesting that opinions were more divided for \f{summary}. Those that ranked \f{summary} highly generally valued it for \ppquote{[making] work so much easier}{P9}, however, some noted how they \ppquote{want to do things [and] feel accomplished}{P30}, goals that were not as well-supported with \f{summary}. Fifteen (47\%) placed \f{prompt} within the top 3 as it was \ppquote{simple and easy}{P23} and \ppquote{encouraged curiosity}{P4}. 
In contrast, seventeen (53\%) ranked \f{feedback} within the bottom 3, and 19 (59\%) ranked \f{none} within the bottom 3. Both techniques required \ppquote{too much effort}{P20}, which could be \ppquote{draining and mind-absorbing}{P25}.

\rev{
\subsubsection{Text Similarity}
We hypothesized that techniques with more human involvement could discourage readers from taking near-verbatim notes \cite{bretzing1979processing}. 
To determine how similar the text typed by the participant was to the selected text, we used Google's Universal Sentence Encoder \cite{cer2018universalSentenceEncoder} to calculate semantic \m{Text Similarity} (0-1 range where 1 means identical).\footnote{\rev{Note that we also calculated \m{Text Similarity} using typed text concatenated with LLM-generated questions and responses for \f{question} and \f{prompt} and found similar results.}} As \f{summary} and \f{blank} do not require the reader to write any text, we compare the selected text to the text that was generated by the LLM. 
Overall, our results suggested that techniques that required writing from the participants were associated with lower \m{Text Similarity} than those that did not. A significant effect of \f{condition} on \m{Text Similarity} and post hoc tests showed that \f{summary} (\median{.78}, \iqr{.14}) and \f{blank} (\median{.8}, \iqr{.15}) had higher \m{Text Similarity} scores than \f{feedback} (\median{.62}, \iqr{.28}), \f{question} (\median{.49}, \iqr{.28}), \f{prompt} (\median{.49}, \iqr{.3}), and \f{none} (\median{.58}, \iqr{.23}). 

}

\rev{
\subsection{Summary}
Our results only suggested marginal differences in reading comprehension, even though many techniques were more mentally demanding and effortful. Participants generally preferred techniques with more AI involvement, even though these techniques were associated with less psychological ownership.
}

\section{Discussion}
\rev{
We summarize principle findings and introduce design implications for each design parameter, then consider how aspects of AI margin notes could transfer to other contexts and acknowledge limitations in our approach. 
}

\rev{
\subsection{Integration}
\emph{Participants had a strong preference for AI margin notes} citing several perceived benefits associated with increased integration: (1) convenient reference to specific concepts in the text when formulating prompts; (2) no need to shift attention to a separate chat interface, and (3) quick access to prior LLM responses.

It is well-known that natural language prompts must be explicit and specific \cite{DirectGPT, DangDiegeticPrompting2023}, but people find this challenging to do \cite{DangDiegeticPrompting2023, MetacognitiveDemandsPrompting2024}. 
Reading and note-taking are already cognitively-demanding activities \cite{piolat2005cognitive}, so the additional task of formulating explicit and specific prompts may be too much \cite{grund2024learning}. 
However, using deictic language is often easier than fully articulated descriptions \cite{ReMap2020, bolt1980put}, which likely applies to prompts as well. 
For example, DirectGPT \cite{DirectGPT} encouraged prompts with deictic language by using direct manipulation to point at deictic references. 
AI margin notes also encourage deictic language by associating each prompt with specific text selected using a standard direct manipulation interaction. 
We believe this frees users to focus more on metacognitive tasks like identifying parts of the document where they require help \cite{MetacognitiveDemandsPrompting2024}, rather than the nuances of language \cite{WhyJohnnyCantPrompt2023}.

\subsubsection*{Design Implications}
The primary implication is to \textbf{adopt an integrated AI margin note approach within document reader software}.
A secondary design implication is to \textbf{devise ways to prompt with more deictic language, less interface switching, and more ways of retrieving previous responses}. For example, allowing readers to place prompts and responses at arbitrary locations independent of text selection, like the `sticky note' feature in most document reader software.
}

\rev{
\subsection{Selection Automation}
\emph{Selecting text manually was slower and required more effort}, but some participants noted this encouraged them to read the text more. Furthermore, \emph{manual selection increased psychological ownership with participants feeling like they were more effective}. This was supported by shorter manual text selections that were placed more uniformly throughout the document. Together, these effects likely contributed to the \emph{strong preferences participants had for manually-created AI margin notes}, with most describing how they \emph{valued the increased control enabled with manual text selections}.

When considering selection automation and human and AI involvement, we observe similar results regarding duration, effort, and psychological ownership. Yet user preferences diverged, with participants preferring more manual control over text selections but more AI involvement within the actual comment. Together, these findings provide additional insight into factors that are more important to users when reading and taking notes.
Based on prior work on text highlighting \cite{yue2015highlighting, joshi2024constrained}, we hypothesize that manual selection forces the reader to identify text they wish to learn more about, which requires them to recognize gaps in their own knowledge or understanding of the text \cite{MetacognitiveDemandsPrompting2024}. An LLM does not know what these gaps are, making it less capable of automatically positioning AI margin notes in ways that align with the reader's needs. Identifying the right text to select is especially important, as the selected text also acts as contextual information for content that is produced and displayed within the comment. As such, \emph{selection automation may be a task that is seen to be less compatible with more automation and AI involvement}. However, once text is manually selected, there is more potential for AI involvement in the creation of the margin note itself.

\subsubsection*{Design Implications}
The primary implication is to \textbf{prioritize manual text selection when creating AI margin notes}. A secondary implication is to \textbf{devise ways of encouraging more control over automatically-placed AI margin notes} when such capabilities are necessary.
For example, readers often struggle to identify the most important information \cite{fowler1974effectiveness}, and systems like Paper Plain \cite{AugustPaperPlain2023} suggest that automatically-generated summaries can help readers understand complex documents. Automatically-placed AI margin notes may have a similar effect, especially when designed in ways that give users more control. One idea is presenting automatically-placed AI margin notes as suggested comments that the reader must manually ``confirm'' to save. This may improve feelings of psychological ownership \cite{Lehmann2022Suggestions}, ensure that the AI margin notes are placed more consistently throughout the document, and encourage users to read the text more closely.

}

\rev{
\subsection{Human and AI Involvement}
}
Our results suggest that \emph{AI margin note techniques have their own strengths and weaknesses}. For example, techniques with more human involvement, like receiving feedback about written text, required more mental demand and effort, but were associated with higher feelings of psychological ownership.  
Yet, participants preferred techniques with more AI involvement, even fully automated summary notes. 
\rev{This contrasts with the desire to manually select text, but it aligns with Kreijkes et al.'s \cite{kreijkes2025effects} findings for LLM-assisted note-taking while reading. In a different note-taking context involving a live video lecture, Chen et al. \cite{ChenAIAssistanceDilemma2025} compared using an LLM to automatically organize generated summaries and transcripts to manual organization by participants, and also found participants preferred more AI involvement. 

However, both Kreijkes et al. \cite{kreijkes2025effects} and Chen et al. \cite{ChenAIAssistanceDilemma2025} also found that note-taking techniques with more human involvement increased comprehension.}
This aligns with work suggesting that increased cognitive engagement can increase learning \cite{bjork1994memory, bjork2011making, bjork1994institutional}, so we are surprised that more cognitively demanding AI margin note techniques were not associated with higher reading comprehension scores. 
To confirm this was not due to our experimental protocol, we conducted other exploratory experiments with different participants and key variations.
We ran experiments with 24 hour gaps between the reading and testing stages to ensure that the documents were not simply remembered in short-term memory, but we did not observe any significant differences. Participants self-declared their knowledge of the topics discussed in each document, so we tried omitting participants with higher background knowledge, and found similar results. Another possibility is a ceiling effect due to a smaller range of possible comprehension scores (0-6). To rule this out, we conducted another experiment using documents and questions from Guidroz et al. \cite{GoogleGuidroz2025llm},\footnote{We emailed the authors and received permission to reuse their materials.} with more granular (0-12) scores. These documents were also at a more difficult university graduate level. Yet again, we found similar results, with no significant differences in comprehension between techniques.

\subsubsection*{Design Implications}
\emph{We believe that less cognitively engaging techniques with more AI involvement, like receiving generated summaries, may not be as detrimental to reading comprehension as one may assume}. \rev{However, our results did suggest that summaries, which featured no human involvement, were not as preferred as other techniques with a little more human involvement, suggesting that readers do not want to offload all responsibilities to AI. Therefore, our primary implication for design is to \textbf{prioritize AI margin note techniques that balance human and AI involvement}. Given the wide range of perceived benefits and trade-offs of each technique, a secondary design implication is to \textbf{provide multiple options of AI margin note techniques that vary in human and AI involvement} to better suit reader preferences and goals. This idea of human and AI involvement could also extend to chat-based interfaces, for example, through techniques that scaffold prompts to elicit more input from the user \cite{MetacognitiveDemandsPrompting2024}.}

\subsection{Other AI Margin Note Designs}

Receiving generated summaries with fill-in-the-blank exercises was the most preferred technique. It was perceived as not too mentally demanding or effortful, still associated with moderate feelings of psychological ownership, and participants felt like they performed well when creating them. \rev{An exciting avenue for future work is to explore other AI margin note designs that focus on these aspects. For example, using digital ink and sketches \cite{romat2019spaceink} to prompt LLMs \cite{yen2025codeshapingiterativecode}. Interactive visual elements, like simulations \cite{AugmentedPhysics} and charts \cite{masson2023charagraph} could be generated by an LLM and integrated as AI margin notes. For example, if the reader selects text that describes different parts of flowering plants, a diagram could be generated with fill-in-the-blank labels for different parts of a flower.
}

\rev{
\subsection{Adapting AI Margin Notes to Other Contexts}

Beyond document reader software, integrating prompting with text selections could prove to be useful in other textual domains. Consider a code editor like VS Code, where code can be selected and an in-line prompt triggered. The explicit text context and ability to use deictic language is similar to AI margin notes, but the prompts and responses are moved to the side chat panel, or not saved at all. Adopting the AI margin note convention of persisting the note in a margin near the associated text (e.g., in the gutter with line numbers) could make prompts easier to reuse and make AI use more transparent for collaborators.

The AI margin note approach could be applied to domains other than text. Clicking on a UI element and prompting using deictic language could be the foundation for general software help-seeking \cite{ReMap2020}. 
This could be automated based on past user behaviour, where multiple AI notes are overlaid on an application interface, each pointing to a part of the UI that typically requires explanation.  
}

\subsection{Limitations}
Our results, especially results related to reading comprehension, may not hold after extended long-term use. For example, LLMs may hallucinate and produce incorrect AI margin notes, and people may come to over-rely by studying incorrect notes \cite{wang2025effect}. Techniques with low human involvement may make students `lazy' over time and impact their ability to take notes in situations where LLMs are unavailable. \rev{Assessing the impact of AI margin note techniques in a wider range of educational settings is an important direction for future work.}

Some aspects of our experimental design may be lacking in ecological validity. Notably, we chose to keep the number of AI margin notes and chat responses constant across techniques, and participants could only interact with one technique at a time. \rev{We made these decisions for increased control. In a real system, these restrictions would not exist:} users could create as many AI margin notes and in whatever way they wish.

\rev{
Our idea of integrating prompting into comments, which are inherently smaller in size, means that there are limits of how much content can be displayed and how many rounds of interactions are possible: with too much content, each AI margin note could become too `chat-like,' nullifying any perceived benefits over chat-based interfaces. Future work could explore ways to adapt or extend AI margin notes when considering this space constraint.
}

\section{Conclusion}
We propose and explore the design of ``AI margin notes'' that leverage the commenting feature of document reader software to provide LLM capabilities in a way that is more integrated into document text. Three experiments evaluated variations from different design parameters: integration, selection automation, and human and AI involvement, and overall, participants valued having integrated AI margin notes and creating them manually. AI margin note techniques that involved the human and AI to different degrees were valued for different reasons, suggesting that document reader software should provide multiple variations to support different user goals and preferences.
Our work adds more evidence that chat-based interfaces are not the only way of interacting with LLMs, and that increased integration with document text is beneficial, especially when they are created manually and the trade-offs of human and AI involvement are considered.

\begin{acks}
This work was made possible by 
NSERC Discovery Grant 2024-03827.  
\end{acks}

\bibliographystyle{lib-acm/ACM-Reference-Format}
\bibliography{main_acm.bib}

\appendix
\makeatother
\clearpage
\newpage
\onecolumn
\flushbottom
\section{Appendix}
\renewcommand\thefigure{\thesection.\arabic{figure}}
\renewcommand\thetable{\thesection.\arabic{table}}
\setcounter{figure}{0}
\setcounter{table}{0}

\begin{table*}[h]
    \centering
    \caption{Experiment 1 statistical test results.}
    \small %
\begin{tabular}{l|r|r|r}
\toprule
Measure & $W$ & $p$ & $RBC$ \\
\midrule
\m{Reading Comprehension} & 80 & .81 & .06 \\
\m{Duration} & 148 & .50 & .16 \\
\m{Psychological Ownership} & 82.5 & .40 & .21 \\
\m{Mental Demand} & 149 & .72 & .08 \\
\m{Physical Demand} & 76 & .68 & .11 \\
\m{Temporal Demand} & 74.5 & .63 & .13\\
\m{Performance} & 70.5 & .78 & .08 \\
\m{Effort} & 105.5 & .32 & .24 \\
\m{Frustration} & 57 & .12 & .40\\
\m{Frequency of Use} & 77.5 & .07 & .44\\
\bottomrule
\end{tabular}
    \label{tab:exp1Stats}
\end{table*}

\begin{table*}[h]
    \centering
    \caption{Experiment 2 statistical test results.}
    \small %
\begin{tabular}{l|r|r|rr}
\toprule
Measure & $W$ & $p$ & $RBC$ \\
\midrule
\m{Reading Comprehension} & 88.5 & .20 & .30 \\
\m{Duration} & 102 & .006 & .56 & ** \\
\m{Psychological Ownership} & 9 & < .001 & .95 & *** \\
\m{Mental Demand} & 158.5 & .46 & .16 \\
\m{Physical Demand} & 31.5 & .10 & .48 \\
\m{Temporal Demand} & 100 & .25 & .28\\
\m{Performance} & 52.5 & .02 & .58 & * \\
\m{Effort} & 84.5 & .004 & .61 & ** \\
\m{Frustration} & 77.5 & .19 & .33\\
\m{Frequency of Use} & 63.5 & .004 & .64 & **\\
\m{Selection Word Count} & 1320 & .003 & .36 & **\\
\bottomrule
\end{tabular}
    \label{tab:exp2Stats}
\end{table*}

\begin{table*}[h]
    \centering
    \caption{Experiment 3 statistical test results.}
    
\small %
\begin{tabular}{rl|r|r|rrl}
\toprule
&Measure & $Q$ & $p$ & $W$ \\
\midrule
&\m{Reading Comprehension} & 10.87 & .05 & .07 \\
(a) & \m{Duration} & 64.64 & < .001 & .40 & *** \\
(b) & \m{Psychological Ownership} & 74.76 & < .001 & .47 & *** \\
(c) & \m{Mental Demand} & 57.83 & < .001 & .36 & *** \\
& \m{Physical Demand} & 20.92 & < .001 & .13 & *** & \emph{post hocs were n.s.} \\
& \m{Temporal Demand} & 4.45 & .49 & .03\\
(d) & \m{Performance} & 16.82 & .005 & .10 & ** \\
(e) & \m{Effort} & 55.98 & < .001 & .35 & *** \\
& \m{Frustration} & 5.39 & .37 & .03\\
& \m{Frequency of Use} & 9.19 & .10 & .06 \\
(f) & \m{Text Similarity} & 92.96 & < .001 & .58 & ***\\
\bottomrule
\end{tabular}

\newcommand{\tabspacelg}{\vspace{0.8em}}
\newcommand{\tabspacesm}{\vspace{0.25em}}

\begin{tabular}{llrrrrrrrrrrrr}
\\
\toprule
&& \multicolumn{2}{l}{\tabspacesm (a) \m{Duration}} & \multicolumn{2}{l}{\tabspacesm (b) \m{P. Ownership}} & \multicolumn{2}{l}{\tabspacesm (c) \m{Mental Demand}} & \multicolumn{2}{l}{\tabspacesm (d) \m{Performance}} & \multicolumn{2}{l}{\tabspacesm (e) \m{Effort}} & \multicolumn{2}{l}{\tabspacesm (f) \m{Text Similarity}}\\
\multicolumn{2}{l}{\textit{comparisons}} & \multicolumn{2}{l}{\textit{p-value}} & \multicolumn{2}{l}{\textit{p-value}} & \multicolumn{2}{l}{\textit{p-value}} & \multicolumn{2}{l}{\textit{p-value}} & \multicolumn{2}{l}{\textit{p-value}} & \multicolumn{2}{l}{\textit{p-value}}\\
\f{none} & \f{feedback} & .02 & * & .24 & & .52 & & 1 & &.91 & & .07\\
\f{none} & \f{question} & .56 & & < .001 & *** & .06 & & 1 && .004 & ** & .09\\
\f{none} & \f{prompt} & .45 & & .003 & ** & .02 & * & 1 && .01 & * & < .001 & ***\\
\f{none} & \f{blank} & .003 & ** & < .001 & *** & < .001 & *** & .005 & ** & < .001 & *** & < .001 & ***\\
\f{none} & \f{summary} & < .001 & *** & < .001 & *** & < .001 & *** & 1 & & < .001 & *** & < .001 & ***\\
\f{feedback} & \f{question} & .18 &  & .002 & ** & .007 & ** & 1 && .01 & * & .002 & **\\
\f{feedback} & \f{prompt} & .04 & * & .003 & ** & .008 & ** & 1 && .01 & * & < .001 & ***\\
\f{feedback} & \f{blank} & < .001 & *** & < .001 & *** & < .001 & *** & .03 & * & .001 & ** & < .001 & ***\\
\f{feedback} & \f{summary} & < .001 & *** & < .001 & *** &  < .001 & *** & 1 && < .001 & *** & < .001 & ***\\
\f{question} & \f{prompt} & .19 & & .44 & & .52 & & 1 & & .91 & & .007 & **\\
\f{question} & \f{blank} & .001 & ** & .004 & ** & .05 & & .08 & & .03 & * & < .001 & ***\\
\f{question} & \f{summary} & < .001 & *** & < .001 & *** & < .001 & *** & 1 & &  .004 & ** & < .001 & ***\\
\f{prompt} & \f{blank} & .03 & * & .24 & & .15 & & .07 && .11 & & < .001 & ***\\
\f{prompt} & \f{summary} & .002 & ** & < .001 & *** & .02 & * & 1 && .02 & * & < .001 & ***\\
\f{blank} & \f{summary} & .14 & & .003 & ** & .15 & & .04 & * & .04 & * & .05 \\
\bottomrule
\end{tabular}

    \label{tab:exp3Stats}
\end{table*}

\end{document}